\documentstyle[draft,psfig]{mn}

\newif\ifAMStwofonts

\ifoldfss
  \ifCUPmtlplainloaded \else
    \NewTextAlphabet{textbfit} {cmbxti10} {}
    \NewTextAlphabet{textbfss} {cmssbx10} {}
    \NewMathAlphabet{mathbfit} {cmbxti10} {} 
    \NewMathAlphabet{mathbfss} {cmssbx10} {} 
  \fi
  \ifAMStwofonts
    \ifCUPmtlplainloaded \else
      \NewSymbolFont{upmath} {eurm10}
      \NewSymbolFont{AMSa} {msam10}
      \NewMathSymbol{\upi}     {0}{upmath}{19}
      \NewMathSymbol{\umu}     {0}{upmath}{16}
      \NewMathSymbol{\upartial}{0}{upmath}{40}
      \NewMathSymbol{\leqslant}{3}{AMSa}{36}
      \NewMathSymbol{\geqslant}{3}{AMSa}{3E}

      \let\leq=\leqslant \let\le=\leqslant
      \let\geq=\geqslant 
    \fi
  \fi
\fi 

\ifnfssone
  \newmathalphabet{\mathit}
  \addtoversion{normal}{\mathit}{cmr}{m}{it}
  \addtoversion{bold}{\mathit}{cmr}{bx}{it}
  \newmathalphabet{\mathbfit} 
  \addtoversion{normal}{\mathbfit}{cmr}{bx}{it}
  \addtoversion{bold}{\mathbfit}{cmr}{bx}{it}
  \newmathalphabet{\mathbfss} 
  \addtoversion{normal}{\mathbfss}{cmss}{bx}{n}
  \addtoversion{bold}{\mathbfss}{cmss}{bx}{n}
  \ifAMStwofonts
    \ifCUPmtlplainloaded \else
      \UseAMStwoboldmath
      \makeatletter
      \new@mathgroup\upmath@group
      \define@mathgroup\mv@normal\upmath@group{eur}{m}{n}
      \define@mathgroup\mv@bold\upmath@group{eur}{b}{n}
      \edef\UPM{\hexnumber\upmath@group}
      \new@mathgroup\amsa@group
      \define@mathgroup\mv@normal\amsa@group{msa}{m}{n}
      \define@mathgroup\mv@bold\amsa@group{msa}{m}{n}
      \edef\AMSa{\hexnumber\amsa@group}
      \makeatother
      \mathchardef\upi="0\UPM19
      \mathchardef\umu="0\UPM16
      \mathchardef\upartial="0\UPM40
      \mathchardef\leqslant="3\AMSa36
      \mathchardef\geqslant="3\AMSa3E

      \let\leq=\leqslant \let\le=\leqslant
      \let\geq=\geqslant 
    \fi
  \fi
\fi 

\ifnfsstwo
  \DeclareMathAlphabet{\mathbfit}{OT1}{cmr}{bx}{it}
  \SetMathAlphabet\mathbfit{bold}{OT1}{cmr}{bx}{it}
  \DeclareMathAlphabet{\mathbfss}{OT1}{cmss}{bx}{n}
  \SetMathAlphabet\mathbfss{bold}{OT1}{cmss}{bx}{n}
  \ifAMStwofonts
    \ifCUPmtlplainloaded \else
      \DeclareSymbolFont{UPM}{U}{eur}{m}{n}
      \SetSymbolFont{UPM}{bold}{U}{eur}{b}{n}
      \DeclareSymbolFont{AMSa}{U}{msa}{m}{n}
      \DeclareMathSymbol{\upi}{0}{UPM}{"19}
      \DeclareMathSymbol{\umu}{0}{UPM}{"16}
      \DeclareMathSymbol{\upartial}{0}{UPM}{"40}
      \DeclareMathSymbol{\leqslant}{3}{AMSa}{"36}
      \DeclareMathSymbol{\geqslant}{3}{AMSa}{"3E}

      \let\leq=\leqslant \let\le=\leqslant
      \let\geq=\geqslant 
    \fi
  \fi
\fi 

\ifCUPmtlplainloaded \else
  \ifAMStwofonts \else 
    \def\upi{\pi}
    \def\umu{\mu}
    \def\upartial{\partial}
  \fi
\fi

\title{Frequentist Estimation of Cosmological Parameters
from the MAXIMA-1 Cosmic Microwave Background Anisotropy Data}
\author[M.E. Abroe et al.]
       {M.~E.~Abroe$^{1}$\thanks{mabroe@physics.umn.edu},
  A.~Balbi$^{2,3}$,
  J.~Borrill$^{4}$,
  E.~F.~Bunn$^{5}$,
  S.~Hanany$^{1}$,
  P.~G.~Ferreira$^{6}$,\newauthor
  A.~H.~Jaffe$^{7,8}$, 
  A.~T.~Lee$^{3,9}$,
  K.~A.~Olive$^{1,10}$,
  B.~Rabii$^{9}$,
  P.~L.~Richards$^{8,9}$,\newauthor
  G.~F.~Smoot$^{3,8,9}$,
  R.~Stompor$^{8,11}$,
  C.~D.~Winant$^{9}$,
  J.~H.~P.~Wu$^{7,12}$\\
   $^1$School of Physics and Astronomy, 116 Church St. S.E., University of
  Minnesota, Minneapolis, MN55455, USA\\
   $^2$Dipartimento di Fisica, Universit\`a Tor Vergata, Roma,
Via della Ricerca Scientifica, I-00133,
  Roma, Italy\\
 $^3$Lawrence Berkeley National Laboratory, 1 Cyclotron Road,
 Berkeley, CA94720, USA\\
$^4$National Energy Research Scientific Computing Center,
  Lawrence Berkeley National Laboratory, Berkeley, CA94720, USA\\
$^{5}$Physics Department, St. Cloud State 
University, St. Cloud, MN56301, USA\\
$^{6}$Astrophysics, University of Oxford, NAPL, Keble Road, Oxford, OX1 3RH,
 UK\\
$^7$Dept. of Astronomy, 601 Campbell Hall, University of California,
  Berkeley, CA94720-3411, USA\\
 $^8$Space Sciences Laboratory, University of California,
  Berkeley, CA94720, USA\\
$^{9}$Dept. of Physics, University of California,
  Berkeley, CA94720-7300, USA\\
$^{10}$Theoretical Physics Institute, University of
  Minnesota, Minneapolis, MN55455, USA \\
$^{11}$Copernicus Astronomical Center, Bartycka 18, 00-716 Warszawa, Poland\\
$^{12}$Department of Physics, National Taiwan University, Taipei 106, Taiwan
}

\pagerange{\pageref{firstpage}--\pageref{lastpage}}
\pubyear{2001}

\begin{document}

\maketitle

\label{firstpage}

\begin{abstract}
We use a frequentist statistical approach to set confidence intervals
on the values of cosmological parameters using the MAXIMA-1 and COBE
measurements of the angular power spectrum of the cosmic microwave
background.  We define a
$\Delta \chi^{2}$ statistic, simulate the measurements of
MAXIMA-1 and COBE, determine
the probability distribution of the statistic, and use it
and the data
to set confidence intervals on several cosmological parameters.
We compare the frequentist confidence intervals to
Bayesian credible regions.
The frequentist and Bayesian approaches give best estimates for 
the parameters that agree within 15\%, and confidence
interval-widths that agree within 30\%.
The results also suggest that a frequentist 
analysis gives slightly broader confidence intervals than a Bayesian 
analysis. 
The frequentist analysis gives values of $\Omega=0.89{+0.26\atop -0.19}$, 
$\Omega_{\rm B}h^2=0.026{+0.020\atop -0.011}$ and $n=1.02{+0.31\atop -0.10}$,
and the Bayesian analysis gives values of $\Omega=0.98{+0.14\atop -0.19}$, 
$\Omega_{\rm B}h^2=0.029
{+0.015\atop -0.010}$, and $n=1.18{+0.10
\atop -0.23}$, all at the 95\% confidence level.
\end{abstract}


\section{Introduction}

The angular power spectrum of the 
temperature anisotropy of the cosmic microwave background (CMB)
depends on parameters that determine the initial and
evolutionary properties of our universe. An accurate measurement of
the power spectrum can provide strong constraints
on these parameters.  During the last few years
several experiments have clearly measured
the first peak in the power spectrum \cite{shaul,de,toco}, 
at an angular scale of about
0.5 degree, and provided evidence for harmonic peaks at smaller angular scales
\cite{lee,boom,dasi}. Detailed analyses have yielded the values of the
total energy density of the universe, the baryon density, the
spectral index of primordial fluctuations, and other parameters with
unprecedented accuracy \cite{boom,pryke,douspis,tegmark,radek}.

Most CMB analyses to date have used a 
Bayesian statistical approach to estimate
the values of the cosmological parameters. 
The results of these analyses can have considerable dependence on the
priors assumed, (e.g., Bunn et al. 1994; Lange et al. 2001; 
Jaffe et al. 2001), and it is 
therefore instructive to attempt an estimate of the
cosmological parameters that is independent of priors. 

Bayesian and frequentist methods for setting limits on parameters
involve quite different fundamental assumptions.  In the Bayesian
approach one attempts to determine the probability distribution of the
parameters given the observed data.  A Bayesian credible region for a
parameter is a range of parameter values that encloses a fixed amount
of this probability.  In the frequentist approach, on the other hand,
one computes the probability distribution of the data as a function
of the parameters.  A parameter value is ruled out if the probability
of getting the observed data given this parameter is low.  Because the
questions asked in the two approaches are quite different, there is no
guarantee that uncertainty intervals obtained by the two methods will
coincide.

Frequentist analyses quantify the probability distribution of the data
in terms of a statistic that quantifies the goodness-of-fit of a model
to the data. The maximum-likelihood estimator $\chi^{2}$ is probably
the most widely used statistic. When the data are Gaussian-distributed
and the model depends linearly on the parameters, the $\chi^{2}$
statistic is $\chi^{2}$-distributed and standard $\chi^{2}$ tables are
used to determine confidence intervals \cite{nr}.

It has become common to compare CMB data to theoretical
predictions via the angular power spectrum, which depends on a number
of cosmological parameters. The data points are usually the most
likely levels of temperature fluctuation power $C_{\ell}$ within
certain bands $\Delta \ell$ of spherical harmonic multipoles.
However, the band powers $C_{\ell}$ are not Gaussian distributed
\cite{bjk}, and the theoretical angular power spectrum does not depend
linearly on the cosmological parameters.  Thus a $\chi^{2}$ statistic
may not be $\chi^{2}$-distributed.  Furthermore, the complicated
probability distribution of the data points and the dependence of the
theoretical predictions on the parameter values make the analytic
calculation of the probability distribution of $\chi^2$ impossible.
Thus, there is no guidance on how to set frequentist confidence
intervals. 

In the past, G\'orski, Stompor \& Juszkiewicz (1993) used a frequentist
analysis to assess the probability of a standard CDM cosmological model
given the data from the UCSB South Pole and COBE experiments (see also
Stompor \& G\'orski, 1994).
More recently,
Padmanabhan \& Sethi (2000), and Griffiths, Silk, \& Zaroubi (2001) 
used a frequentist approach
to determine confidence intervals on several cosmological
parameters.
These recent analyses (implicitly) assume that the band power $C_{\ell}$ 
is Gaussian distributed and that the
cosmological model is linear in the cosmological parameters, and thus
that standard $\chi^{2}$ values can be used to set confidence
intervals on various cosmological parameters.  These analyses also do
not account for correlations between between band powers. 
Gawiser (2001) argued that a frequentist analysis is better 
suited than Bayesian for
answering the question of how consistent parameter estimates from CMB
data are with estimates from other astrophysical
measurements. 
A method for estimating the angular power spectrum
which uses frequentist considerations was
presented in Hivon et al. (2001).

In this paper we present a more rigorous approach to frequentist
parameter estimation from CMB data than previous analyses.
We use the data from the COBE \cite{gorski} and 
MAXIMA-1 experiments \cite{shaul} 
and simulations to determine the probability distribution of an
appropriate $\Delta \chi^{2}$ statistic, and use this distribution
to set frequentist confidence intervals on several cosmological
parameters. We compare the frequentist confidence intervals to 
Bayesian credible 
regions obtained using the same data and to the likelihood-maximization
results of Balbi et al. (2000). 

The structure of this paper is as follows: in Section~\ref{sec:data}
we discuss the MAXIMA-1 and COBE data and the database of cosmological
models used in our analysis.  In Section~\ref{sec:chi2} we present
the $\chi^2$ statistic used in our analysis.  Section~\ref{sec:conf}
describes the process of setting frequentist and Bayesian confidence regions on
cosmological parameters.  The results and a discussion are given in
Sections~\ref{sec:results} and~\ref{sec:conclusions}.

\section{Data and Database of cosmological models}
\label{sec:data}

We use the angular power spectrum computed from the 5$'$ MAXIMA-1 CMB
temperature anisotropy map \cite{shaul} and the 4-year COBE
angular power spectrum \cite{gorski}. The MAXIMA-1 and COBE power
spectra have 10 and 28 data points in the range $36\leq\ell\leq785$
and $2\leq\ell\leq35$, respectively. 
Lee et al. (1999) and 
Hanany et al. (2000)
provide more information about the MAXIMA experiment and data. 
Santos et al. (2001) and Wu et al. (2001a) showed that the 
temperature fluctuations in the MAXIMA-1 map are 
consistent with a Gaussian distribution.   
Lee et al. (2001) have recently extended the analysis of the 
data from MAXIMA-1 to smaller angular scales, but these data are not 
used in this paper. 

To perform our analysis we constructed a database of 
330,000 inflationary cosmological models \cite{cmbfast} 
that has the following 
cosmological parameter ranges and resolutions:
\begin{itemize}
\item $\tau= 0.0,0.1,0.2,0.3,0.4,0.5$
\item $\Omega_{\rm B}=0.005,0.01,0.02,0.03,0.04,0.05,0.075,0.10,0.15$
\item $\Omega_{\rm M}=0.05,0.1,0.15,0.2,0.25,0.3,0.35,0.4,0.5,0.6,0.7,$\\
\hspace*{.57in} 
$0.8,0.9,1.0$
\item $\Omega_{\Lambda}=0.0,0.1,0.2,0.3,0.4,0.5,0.6,0.7,0.8,0.9,1.0$
\item $H_0=40,50,60,70,80,90$
\item $n=0.6,0.7,0.8,0.9,1.0,1.1,1.2,1.3,1.4,1.5$
\end{itemize}
The parameter $\tau$ is the optical depth to reionization, $n$ is the
scalar spectral index of the primordial power spectrum, and $H_0$ is
the Hubble parameter in units of $\rm km \,s^{-1} \,Mpc^{-1}$.  The
density parameters $\Omega_{\rm B}$, $\Omega_{\rm M}$, 
and $\Omega_{\Lambda}$
give the ratios of the density of baryons, total matter, and cosmological
constant to the critical density.

\section{The $\chi^2$ Statistic} 
\label{sec:chi2}

To set frequentist confidence intervals we choose the
maximum-likelihood estimator $\chi^{2}$ as a goodness-of-fit
statistic.  We use the $\chi^2$ as defined in equation (39) of Bond,
Jaffe, \& Knox (2001, hereinafter BJK) 

\begin{eqnarray}
\label{eqn:chi2}
\chi^2 & = &\sum^{{\rm }}_{i,j}(Z_i^{d}-Z_i^t)
{\bf\sf M}_{ij}(Z_j^{d}-Z_j^t) + {{(u-1)^2}\over{\sigma_u^2}} \label{line2} 
\end{eqnarray}
\begin{equation}
Z_i^{t}=\ln({\cal N}C_i^{t} + x_i)
\end{equation}
\begin{equation}
Z_i^{d}=\ln(uC_i^{d} + x_i).
\label{eqn:Z}
\end{equation}
The sum in equation~(\ref{eqn:chi2}) includes the COBE and MAXIMA-1 bands. 
The data and theory band powers are denoted as 
$C^d_i$ and $C^t_i$, respectively, ${\bf\sf M}_{ij}$ is the inverse
covariance matrix for the $Z_i^d$ quantities, and ${\cal N}$ is the
normalization of the models to the data [sometimes called
$C_{10}$, e.g. Balbi et al. (2000)]. The 
variable $u$ accounts for the calibration uncertainty
of the MAXIMA-1 data, which is 8\% in 
the power spectrum \cite{shaul}, i.e. $\sigma_{u} = 0.08$. 
For the COBE bands $u$ is defined to be one. 
\begin{figure}
\centerline{\psfig{file=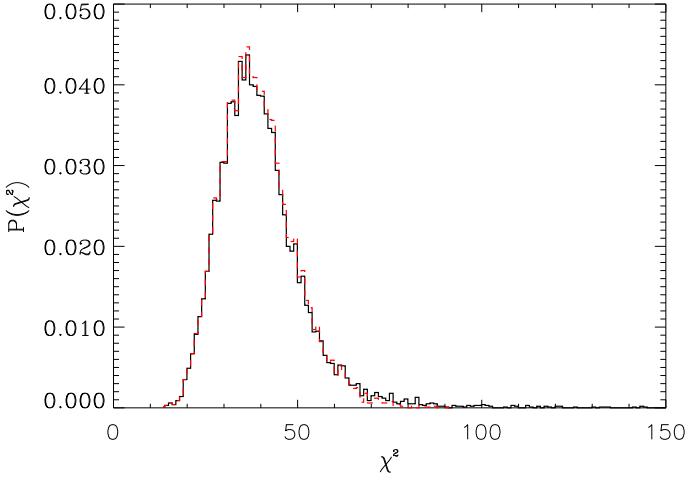,width=3.8in,angle=90}}
\caption{\footnotesize Distribution of $\chi^2$ 
values computed for the 10,000 simulations by 
solving for ${\cal N}$ and $u$ analytically
using the method described in the text (dashed
line), and numerically using Brent's root-finding algorithm 
(solid line).
Both histograms have a bin size of one and
are normalized to integrate to one.}
\label{max_comp}
\end{figure}

Each time a $\chi^2$ is calculated we solve for the normalization
${\cal N}$ and calibration factor $u$ that simultaneously minimize
$\chi^2$.  Because we did not find a closed-form analytical solution
to the minimization of equation~(\ref{eqn:chi2}) with respect to ${\cal
N}$ and $u$, and a numerical minimization would have been
computationally prohibitive, we used the following approximation.  We
assume that equation~(\ref{eqn:chi2}) is well approximated by
\begin{equation}
\label{eqn:chi2_app}
{\chi^2}  \simeq  \sum_{i,j}(uC_i^{d}-{\cal N}C_i^t){\bf\sf F}_{ij}
(uC_j^{d}-{\cal N}C_j^t) 
+{{(u-1)^2}\over{\sigma_u^2}} \label{line2}.
\end{equation}
Where the Fisher matrix ${\bf\sf F}_{ij}$ for the $C_i^d$ quantities,
is related to ${\bf\sf M}_{ij}$ by
\begin{equation}
{\bf\sf M}_{ij}={\bf\sf F}_{ij}(C_i^d+x_i)(C_j^d+x_j). 
\end{equation}
\begin{figure}
\centerline{\psfig{file=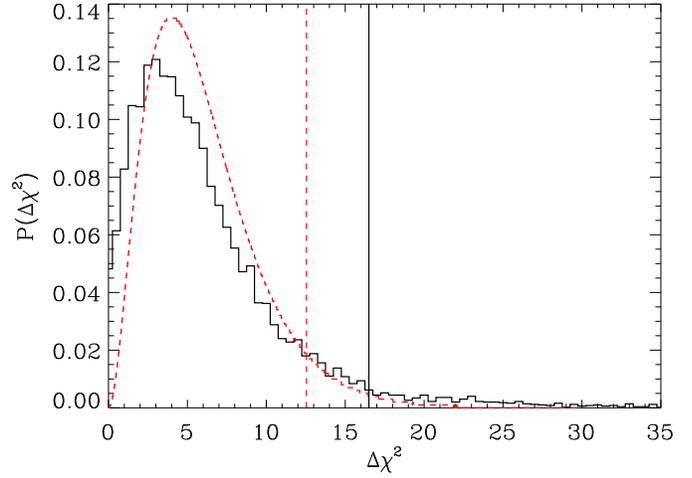,width=3.8in,angle=90}}
\caption{\footnotesize 
\footnotesize The histogram gives the $\Delta \chi^2$
distribution for the entire six-dimensional parameter space
from 10,000 simulations of the COBE and MAXIMA-1 band powers. 
The dashed curve is the standard $\chi^2$ distribution for six 
degrees of freedom.  The vertical solid line (vertical dashed line) 
is $\Delta \chi^2=16.5 \, (12.8)$, which corresponds 
to the 95\% $\Delta \chi^2$ threshold
for the histogram (standard $\chi^2$ distribution).
The histogram has a bin size of 0.5, and is normalized to 
integrate to one.
}
\label{6dim_hist}
\end{figure}
Minimization of equation~(\ref{eqn:chi2_app}) 
with respect to ${\cal N}$ and $u$ gives two coupled equations
which we solve for $u$ by assuming that ${\cal N}=1$.  We then use that value
of $u$ to solve for ${\cal N}$.  We compared this approximate solution to a
rigorous numerical minimization of
equation~(\ref{eqn:chi2}) for 10,000 cases and found an RMS fractional
error of less than 1.5\% (see Figure~\ref{max_comp}).
Once the factors ${\cal N}$ and $u$ have been determined using 
equation~(\ref{eqn:chi2_app}), 
the exact equation~(\ref{eqn:chi2}) is used to find
the value of $\chi^{2}$.

\section[]{Determining Confidence Levels}
\label{sec:conf}

\subsection[]{Frequentist Confidence Intervals}
\label{sec:frequentist_conf}

\begin{figure*}
\centerline{\psfig{file=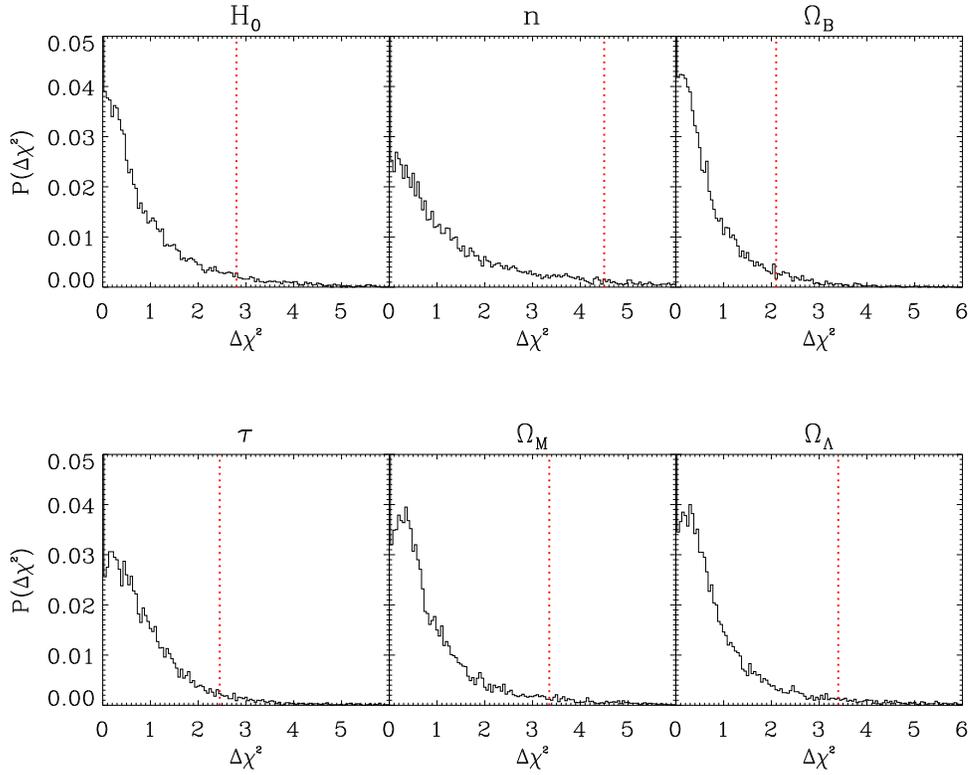,width=6in,angle=90}}
\caption{\footnotesize Simulated one-dimensional  
$\Delta \chi^2$ distributions for all the parameters in the
database. The vertical dotted lines correspond to the 95\% $\Delta
\chi^2$ threshold level; numerical values 
are given in Table 1.  
Each histogram has a bin size of $0.05$ and
is normalized to integrate 
to one. The 95\% threshold for a standard $\chi^2$
distribution with one degree of freedom is
$\chi^2=3.8$.} 
\label{chi2hist}
\end{figure*}

\begin{figure*}
\centerline{\psfig{file=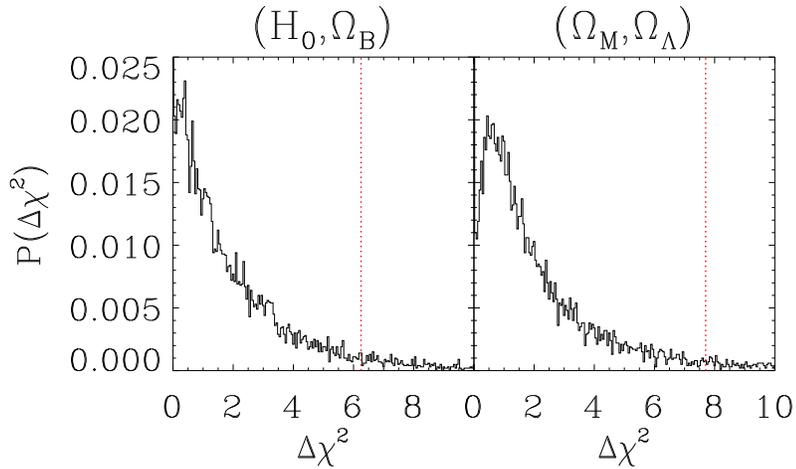,width=4.5in,angle=90}}
\caption{\footnotesize Simulated two-dimensional  
$\Delta \chi^2$ distributions in the $(H_0,\Omega_{\rm B})$ and
$(\Omega_{\rm M},\Omega_{\Lambda})$ planes.  The 
vertical dotted lines correspond to the
95\% $\Delta \chi^2$ cutoff level, which are 
6.25 and 7.70 for $(H_0,\Omega_{\rm B})$ and
$(\Omega_{\rm M},\Omega_{\Lambda})$ respectively.  Each histogram has
a bin size of $0.05$ and is normalized to integrate to one.  The 95\%
threshold for a standard $\chi^{2}$ distribution with two degrees of
freedom is $\chi^2=6$.  }
\label{chi2hist_2}
\end{figure*}

Let ${\bf a}$ denote a vector of parameters in our six-dimensional
parameter space, and ${\bf a}_{\rm true}$ be
the unknown true values of the cosmological parameters
that we are trying to estimate.
By minimizing $\chi^2$ we find that the
best-fitting model 
to the MAXIMA-1 and COBE data has the following parameters ${\bf{a}}_0$:
$(H_0,\Omega_{\rm B},\Omega_{\rm M},\Omega_{\Lambda},n,\tau)
=(60,.075,.7,.2,1,0)$.  This model gives a $\chi^2=36$, which is an
excellent fit to 38 data points.  We define
\begin{equation}
\label{eqn:deltachi2}
\Delta \chi^2({\bf a}) \equiv \chi^2({\bf a})-\chi^2({\bf{a}}_0),
\end{equation} 
where the first term on the right hand side 
is a $\chi^{2}$ of the 
data with a model in the database, and the second is
a $\chi^{2}$ of the data with the best-fitting model. 
To quantify the probability distribution 
of the data as a function of the parameters we choose a 
threshold $\Theta$, and define ${\cal R}$ to be the region in
parameter space such that $\Delta \chi^2 \leq \Theta$.  ${\cal R}$ is
a confidence region at level $\alpha$ if there is a probability
$\alpha$ that ${\cal R}$ contains the true cosmological parameters
${\bf a_{\rm true}}$.  In other words, if many vectors
${\bf{a}}_{0(j)}$ and regions ${\cal R}_{j}$ are generated by
repeating the experiment many times, a fraction $\alpha$ of the
ensemble of ${\cal R}_{j}$ would contain ${\bf a_{\rm true}}$.  
Since the $\Delta\chi^{2}$ statistic may not be $\chi^{2}$ distributed, 
we use simulations to determine its probability distribution
as a function of the cosmological parameters. 
\begin{figure*}
\centerline{\psfig{file=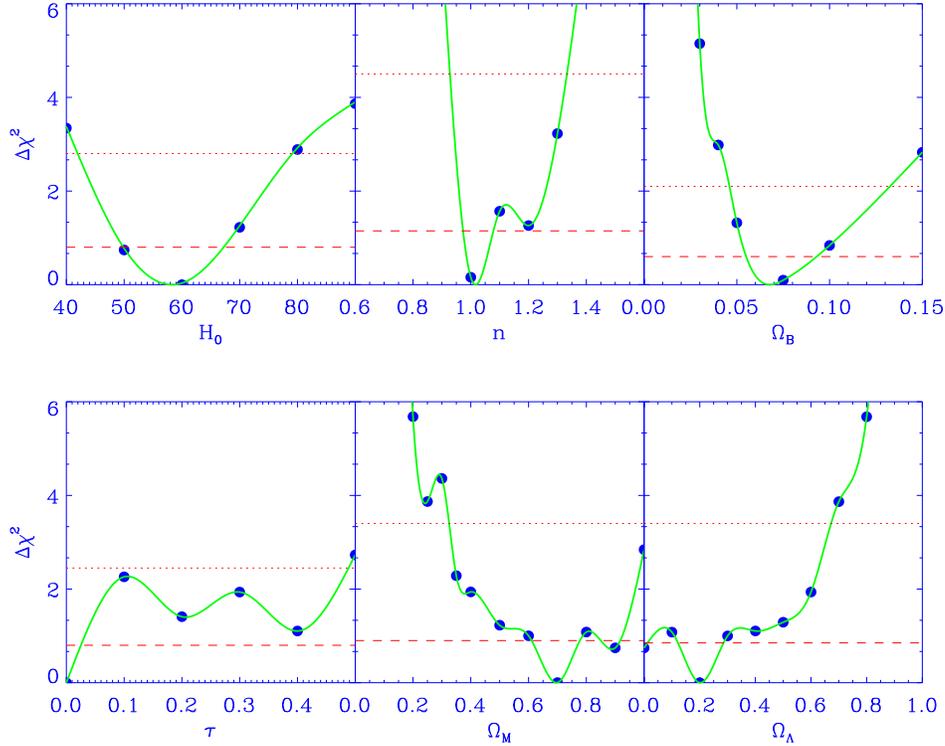,width=5.6in,angle=90}}
\caption{\footnotesize  
$\Delta\chi^2$ calculated 
with the MAXIMA-1 and COBE data as a function of parameter value for
each of the parameters in the database.  Solid circles show grid
points in parameter space, and the solid lines were obtained by
interpolating between grid points.  The parameter values where the
solid line intercepts the dashed (dotted) line corresponds to the
68\% (95\%) frequentist confidence region. 
\label{6chifunc}
}
\end{figure*}
 
The simulations mimic 10,000 independent observations of the CMB
by the MAXIMA-1 and COBE experiments. 
The CMB is assumed to be
characterized by the MAXIMA-1 and COBE best estimate for the
cosmological parameters, ${\bf{a}}_{0}$. 
Applying the equivalent of
equation~(\ref{eqn:deltachi2}) [see equations~(\ref{eqn:chi2j}) 
and (\ref{eqn:chi2one})]
for each of the simulations gives
a set of 10,000 values $\Delta \chi^{2}_{j}$ ($j=1,...,10^4$), 
and by histogramming these
values we associate threshold levels $\Theta$ 
with probabilities $\alpha$.  This
relation between $\Theta$ and $\alpha$ 
is applied to the distribution of $\Delta
\chi^{2}$ that is calculated using equation~(\ref{eqn:deltachi2}) 
to determine a
frequentist confidence interval ${\cal R}$ on the cosmological parameters. 
Note that our procedure assumes that the probability
distribution of $\Delta\chi^2$ around ${\bf{a}}_{0}$ 
closely mimics the probability
distribution of $\Delta\chi^2$ around ${\bf a_{\rm true}}$. 
This is a standard
assumption in frequentist analyses \cite{nr}. The alternative 
approach of
determining the probability distribution around each grid
point in parameter space is computationally prohibitive.

Because finding the best-fitting band powers from a time stream or even a
sky map is computationally expensive \cite{borrill}, we perform
10,000 simulations of the quantities $Z_i^d$, which are 
related to the band powers $C_{\ell}$
as defined in equation~(\ref{eqn:Z}). We assume that the $Z_i^d$ are 
Gaussian-distributed \cite{bjk} and we discuss and justify this assumption in
Appendix~\ref{prob}.

The quantities $Z_{i(j)}^d$, where $j$ denotes one of the 10,000 
simulations, are drawn from two multivariate Gaussian distributions
that represent the MAXIMA-1 (10 data points) and COBE (28 data
points) band powers.  The means of the distributions are the
$Z_{i}^t$ quantities as determined by ${\bf a_{0}}$, and the
covariances are taken from the data.  Each
of the $Z_{i{(j)}}^d$ is thus a vector with 38 elements representing
an independent observation of a universe with a set of cosmological
parameters $\bf{a_{0}}$ .  
We include uncertainty in the calibration and beam-size of the
MAXIMA-1 experiment by multiplying the MAXIMA band-powers by two
Gaussian random variables. The calibration random variable has a mean
of one and standard deviation of 0.08, and the beam-size random
variable has a mean of one and a variance that is $\ell$-dependent
\cite{shaul,wu2}.  For each simulation $j$ the entire database of
cosmological models is searched for the vector of 
parameters ${\bf{a}}_{0(j)}$ which 
minimizes $\chi^2$, and we calculate $\Delta\chi^{2}_{j}$: 
\begin{equation}
\label{eqn:chi2j}
\Delta\chi^{2}_{j} = \chi^{2}_j({\bf a}_0) - \chi^{2}_j({\bf a}_{0(j)}).  
\end{equation}
The first and second terms on the right hand side are the $\chi^{2}$ of
simulation $j$ with the model $\bf{a_{0}}$, and of simulation $j$ with
its best-fitting model, respectively. A normalized
histogram of  
$\Delta\chi^{2}_{j}$ for all 10,000
simulations is shown in Figure~\ref{6dim_hist} and gives the probability
distribution of $\Delta\chi^{2}$ over 
the six-dimensional parameter
space. The 95\% threshold is  $\Delta \chi^2=16.5$, that is, 95\% of the 
probability is contained in the range $0\le \Delta \chi^2 \le 16.5$. 
Figure~\ref{6dim_hist} also shows
a standard $\chi^2-$distribution with
six degrees of freedom and its associated 95\%  threshold level.
The difference between the results of the simulations
and the standard $\chi^2$ distribution is attributed
to the non-linear dependence of the models on the parameters
and minimizing $\chi^2$ over ${\cal N}$ and $u$.

Contour levels in the six-dimensional parameter space that are 
provided by different thresholds of the distribution of $\Delta\chi_j^2$
cannot be used to set confidence intervals on any individual parameter. 
To find a confidence interval for a single parameter $p$ we
compute the probability distribution $\Delta\chi^2_{(p)}$ in the
following way.
We search the database for the model that minimizes the 
$\chi^{2}$ with simulation $j$
under the condition that $p$ is fixed at its value in $\bf{a_{0}}$,
and for the model that minimizes the $\chi^{2}$ with simulation 
$j$ with no restrictions on the parameters.  
We compute
\begin{equation}
\label{eqn:chi2one}
\Delta\chi^{2}_{(p)} = \chi^{2}_j({\bf a}_{(p)}) - 
\chi^{2}_j({\bf a}_{0(j)}), 
\end{equation}
where ${\bf a}_{(p)}$ is the vector of parameters that minimize
$\chi^2$ subject to the constraint that $p$ is fixed. A
histogram of $\Delta\chi^{2}_{(p)}$ provides the necessary
distribution. The one-dimensional distributions for all six parameters in
the database are shown in Figure~\ref{chi2hist}.  Generalization of
this process for finding the probability distribution for any
subset of parameters is straightforward.  The two-dimensional $\Delta
\chi^2$ distributions in the $(H_0,\Omega_{\rm B})$ and
$(\Omega_{\rm M},\Omega_{\Lambda})$ planes are shown in
Figure~\ref{chi2hist_2} and the corresponding 95\% thresholds are
$\Delta \chi^2=6.25$ and $\Delta \chi^2=7.70$, respectively.

\begin{figure}
\centerline{\psfig{file=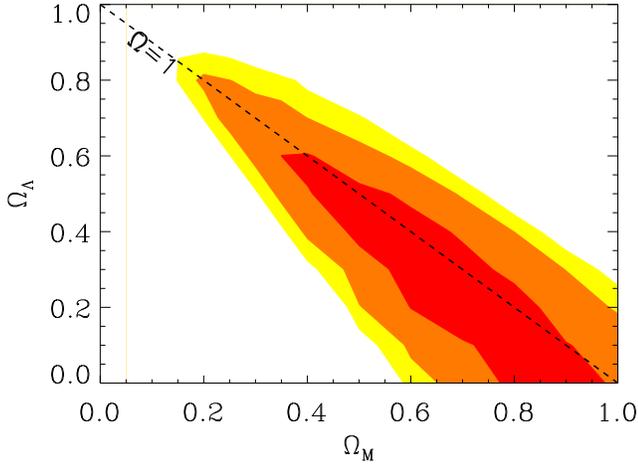,width=3.8in,angle=90}}
\caption{\footnotesize 
Two-dimensional frequentist confidence regions
in the ($\Omega_{\rm M},\Omega_{\Lambda}$) plane.  The 
red, orange, and yellow regions
correspond to the 68\%, 95\%, and 99\% confidence
regions respectively.  The dashed line corresponds
to a flat universe, $\Omega=\Omega_{\rm M}+\Omega_{\Lambda}=1$.}
\label{lm}
\end{figure}

\begin{figure}
\centerline{\psfig{file=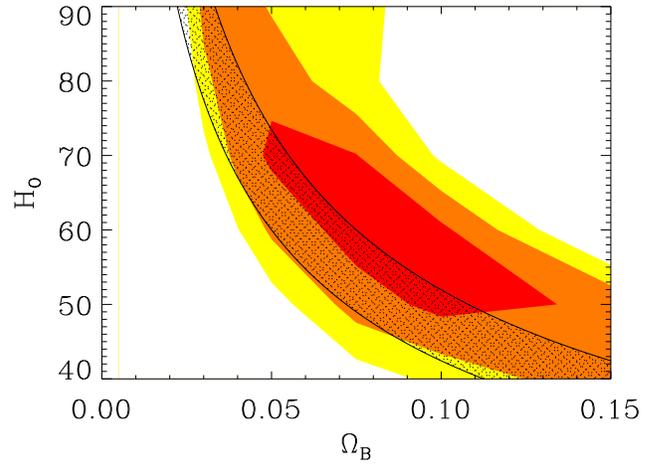,width=3.8in,angle=90}}
\caption{\footnotesize 
Two-dimensional frequentist confidence regions
in the ($H_0,\Omega_{\rm B}$) plane.  The red, orange, and yellow
regions correspond to the 68\%, 95\%, and 99\% confidence
regions respectively.  Standard calculations from big bang 
nucleosynthesis and observations
of ${{D}/{H}}$ predict a 
95\% confidence region of $\Omega_{\rm B}h^2=
0.021{+0.006\atop -0.003}$ \protect\cite{olive}, indicated by the shaded
region.}
\label{hb}
\end{figure}

Using the simulated one- and two-dimensional probability distributions
of $\Delta \chi^{2}$ we set 68\% and 95\% threshold levels on the distribution 
of $\Delta \chi^{2}$ that are calculated using the data and the 
database of models, i.e. the one calculated from  
equation~(\ref{eqn:deltachi2}), and we 
determine corresponding confidence intervals
on the cosmological parameters. Figures~\ref{6chifunc},
\ref{lm}, and \ref{hb} give the  
association between $\Delta \chi^{2}$ and cosmological parameter values
for each of the parameters in the database and in the $(H_0,\Omega_{\rm B})$
and $(\Omega_{\rm M},\Omega_{\Lambda})$ planes.  
\subsection{Bayesian Credible Regions}
\begin{figure*}
\centerline{\psfig{file=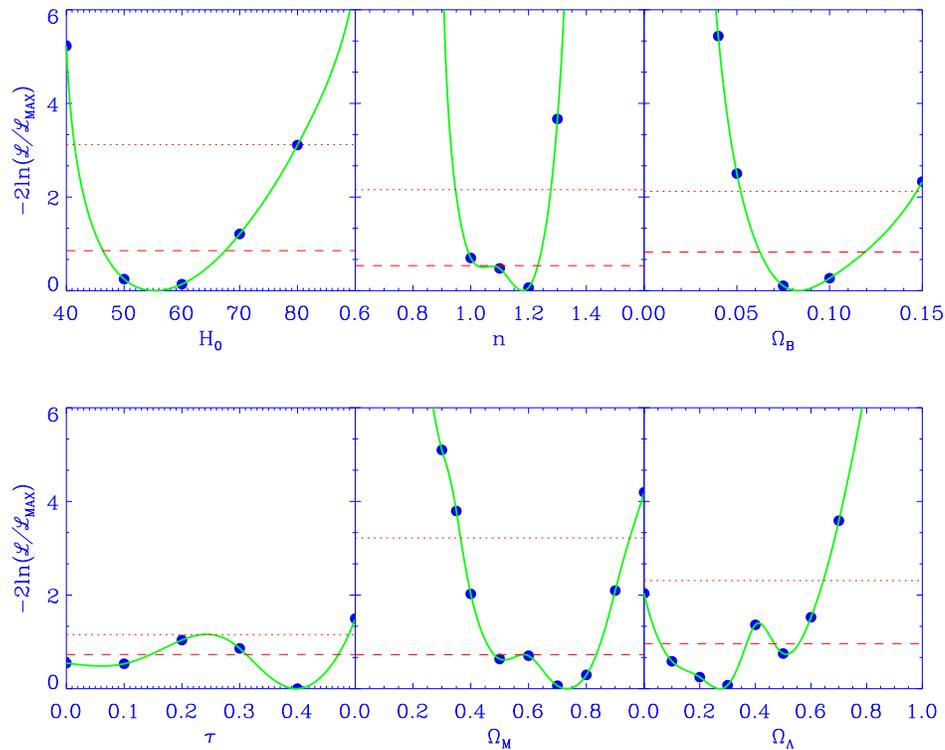,width=5.6in,angle=90}}
\caption{\footnotesize 
Bayesian likelihood functions
for each of the parameters in the database. Solid circles
show grid points in parameter space, while the solid
lines were obtained by interpolating between grid points.
The parameter values where the solid line intercepts
the dashed (dotted) line corresponds to the 68\%
(95\%) Bayesian credible regions.
\label{6bayesfunc}
}
\end{figure*}
\begin{figure*}
\centerline{\psfig{file=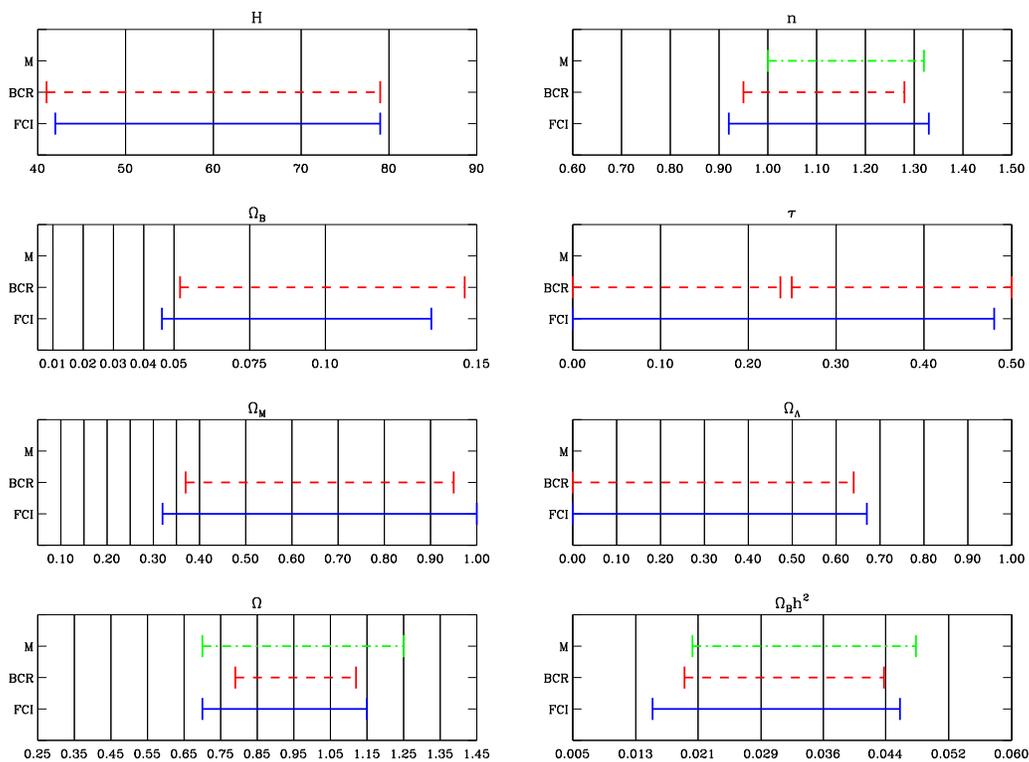,width=5.6in,angle=90}}
\caption{\footnotesize 
A comparison of the 95\% frequentist
confidence intervals (FCI, solid line), Bayesian credible 
regions (BCR, dashed line), and maximization 
regions (M, dashed dot line) for the parameters in the database,
and for $\Omega$ and $\Omega_{\rm B}h^2$.
}  
\label{fig:intervals}
\end{figure*}

According to Bayes's theorem the probability of a model given the
data, the posterior probability,
is proportional to the product of the likelihood ${\cal L}\left({\bf
a}\right)=\exp(-{1\over2}\chi^2({\bf a}))$ 
and a prior probability distribution of the parameters.
If the prior is constant, as we shall assume, then the posterior
probability is directly proportional to the likelihood function.  
To set a Bayesian credible region for any parameter 
or subset of parameters
of interest we calculate the likelihood ${\cal L}({\bf a})$ 
for all models in the database, assume a
flat prior probability distribution for all parameters, and integrate
the likelihood over the remaining parameters. The 95\%
credible region is the region that encloses 95\% of the probability.
The likelihood functions for each parameter in the database are shown
in Figure~\ref{6bayesfunc}.

\section{Results}
\label{sec:results}

The 95\% frequentist confidence intervals and Bayesian credible
regions for each parameter in the database are given in Table 1 and 
Figure~\ref{fig:intervals}.

We found that the optical depth to last scattering $\tau$ was
degenerate with other parameters in the database, mostly with the
spectral index of the primordial power spectrum $n$.  Because of this
degeneracy the 95\% confidence interval of $\tau$ covers nearly the
entire range of values considered, and the 95\% credible region is
disjoint. 

\begin{table*}
\label{tab:conf}
\begin{centering}
\begin{tabular}{|c|c|c|c|c|} \hline
\multicolumn{5}{|c|}{95\% Confidence Regions} \\ \hline
Parameter & $\Delta \chi^2_{(95\%)}$ & frequentist & Bayesian & 
     Balbi et al. \\ \hline \hline
$H_0$              & 2.80 & [42,79] & [41,79] & ---\\ \hline
$n$                & 4.50 & [0.92,1.33] & [0.95,1.28] & [1.00,1.32]\\ \hline
$\Omega_{\rm B}$         & 2.10 & [0.046,0.135] & [0.052,0.146] &--- \\ \hline
$\tau$             & 2.45 & $\leq0.48^{\rm a}$    & ---    &     --- \\ \hline
$\Omega_{\rm M}$         & 3.40 &   $\geq 0.32^{\rm a}$  & [0.37,0.95] & ---\\ \hline
$\Omega_{\Lambda}$ & 3.40 & $\leq0.67^{\rm a}$     &  $\leq0.64^{\rm a}$ &---\\ \hline\hline
$\Omega$           & 5.1 & [0.70,1.15]   & [0.79,1.12]  & [0.70,1.25] \\ \hline
$\Omega_{\rm B}h^2$      & 5.6 & [0.015,0.046] & [0.019,0.044] & [0.020,0.048] \\ \hline
\multicolumn{3}{c}{a: Sets only upper or lower limits on parameter} \\
\end{tabular}
\caption{\footnotesize  A comparison of Bayesian, frequentist 
and maximization 95\%
confidence intervals. The table also gives
the 95\% $\Delta \chi^{2}$ thresholds from the simulations. 
Maximization confidence intervals 
are taken from Balbi et al. (2000); they do not give confidence
intervals for all the parameters. }
\end{centering}
\end{table*}

The 95\% confidence intervals and 
credible regions for $H_0$, $\Omega_{\rm B}$,
$\Omega_{\rm M}$ and $\Omega_{\Lambda}$ include most of the parameter values
because the angular power spectrum of the CMB is 
not very sensitive to any
one of them alone. It is more sensitive to the combination
$\Omega_{\rm B}h^2$, and to the total energy density parameter
$\Omega=\Omega_{\rm M}+\Omega_{\Lambda}$.  To set confidence
intervals and credible regions for these new parameters we
formed all possible combinations of $\Omega$ and $\Omega_{\rm B}h^{2}$ in
our database and binned them in the following 
bins for $\Omega_{\rm B}h^2$:
{\{$0.00500$,$0.0129$,$0.0207$,$0.0286$,$0.0364$,$0.0442$,
$0.0521$,$0.0600$}\}, and for
$\Omega$: \{{$0.25$,$0.35$,$0.45$,$0.55$,$0.65$,$0.75$,
$0.85$,$0.95$,$1.05$,$1.15$,$1.25$,$1.35$,$1.45$}\}.  The center value in each
bin is considered a new database grid point.  We repeat the
process used to find the 95\% $\Delta \chi^2$ threshold for
$\Omega_{\rm B}h^2$ and $\Omega$, treating them as one-dimensional
parameters.  We also calculate the appropriate integrated likelihood
functions.  Table 1 lists the 95\% $\Delta\chi^2$ threshold,
confidence interval, and credible regions for $\Omega$ and
$\Omega_{\rm B}h^2$. 

We determined the frequentist and Bayesian central
values for $n$, $\Omega_{\rm B}h^{2}$ and $\Omega$, the parameters to
which the CMB power spectrum is most sensitive.  In the frequentist 
approach the central value of a parameter is 
the value given by the best-fitting model, and the Bayesian
central value is the maximum marginalized likelihood
parameter value.  
The frequentist and Bayesian
analyses give respectively a value of $\Omega=0.89{+0.26\atop -0.19}$ and
$0.98{+0.14\atop -0.19}$, $\Omega_{\rm B}h^2=0.026{+0.020\atop -0.011}$ and
$0.029{+0.015\atop -0.010}$, and $n=1.02{+0.31\atop -0.10}$ and
$1.18{+0.10\atop -0.23}$, all at the 95\% confidence level.

\section{Discussion}
\label{sec:conclusions}

A comparison of the frequentist confidence intervals and the Bayesian
credible regions is shown in Figure~\ref{fig:intervals}.  
We have also included the
results of Balbi et al. (2000), who set parameter confidence intervals 
using the same data set considered in this paper but
use maximization rather than marginalization of the likelihood function.
In this method the likelihood function for a parameter is determined
by finding the
maximum of the likelihood ${\cal L}({\bf a})$ as a function
of the remaining parameters.
When ${\cal L}({\bf a})$ is Gaussian
maximization and marginalization
are equivalent.

The central values and the widths of
confidence intervals derived from all three methods give
consistent results within about 15\% and
30\% respectively. A closer examination
suggests, however, that frequentist confidence intervals are somewhat
broader than the Bayesian ones.  For five out of eight parameters
($n$, $\Omega_{\rm M}$, $\Omega_{\Lambda}$, $\Omega$,
and $\Omega_{\rm B}h^2$) the
frequentist confidence intervals are somewhat larger than the Bayesian
credible regions, and for $H_0$ the intervals are nearly identical.
For the three parameters for which we have results from all three methods
the Bayesian intervals are either the narrowest ($\Omega_{\rm B}h^{2}$
and $\Omega$) or identical to maximization ($n$).
Also, we find that the 99\% frequentist confidence intervals are
somewhat wider than the 99\% Bayesian credible regions for
every parameter considered except $\Omega_{\rm B}$.

Despite this pattern which suggests that a frequentist analysis gives
broader confidence intervals, it is difficult to claim such a pattern
conclusively.  Furthermore, it is not useful to quantify the pattern exactly
because the difference in confidence interval widths is usually within
one parameter grid point. Much finer gridding and hence a much larger
database would be necessary to claim such a pattern with high confidence.
A larger database would also provide a more accurate determination
of the $\Delta\chi^2$ functions in Figure~\ref{6chifunc}, and
the likehood functions in Figure~\ref{6bayesfunc}.  However, a
larger database would have been computationally prohibitive.

The difference between the confidence interval and credible region for
the baryon density $\Omega_{\rm B} h^2$ is of some interest.  Maximization 
\cite{balbi} and Bayesian \cite{jaffe} analyses 
of the MAXIMA-1 and COBE data gave
consistency between $\Omega_{\rm B}h^{2}$ from CMB measurements and a
value of 0.021 from some determinations of ${{D}/{H}}$ from quasar
absorption regions \cite{bbn} only at the edge of the 95\%
intervals. This was interpreted as a tension between CMB measurements
and either deuterium abundance measurements or calculations of BBN
\cite{teg2,silk,olive2}.  A value of 0.021 for $\Omega_{\rm
B}h^2$ is consistent with the frequentist confidence interval at a
level of 75\%, an agreement at a confidence of just over $1 \sigma$.
Recent analysis of new CMB data is consistent with a value of
$\Omega_{\rm B}h^2=0.021$ within a $1 \sigma$ level
\cite{pryke,boom}.

The comparison between the Bayesian- and frequentist-based
analyses raises the question of whether agreement at the
level observed was in fact expected.
The Bayesian and frequentist approaches to parameter estimation are
conceptually quite different.  A Bayesian asks how
likely a parameter is to take on any particular value, given the
observed data.  A frequentist, on the other hand, asks how likely
the given data set is to have occurred, given a particular set of
parameters.  Since the two questions are completely different, there
is no guarantee that they will yield identical answers in
general.
In certain specific situations Bayesian and frequentist approaches
can be shown to yield the same results.  For example, in the
particular case of Gaussian-distributed data with uniform priors and
linear dependence of the predictions on the parameters, the two
approaches coincide.  However, these hypotheses (particularly the last
one) do not apply to the case we are considering.

Bayesian and frequentist methods also coincide asymptotically, 
i.e. in the limit as the number of independent data points tends to
infinity (Ferguson 1996).  In that limit,
all confidence regions would be small in comparison to the
prior ranges of the parameters, and the Bayesian prior-dependence
would become negligible.  CMB data are clearly not yet in this
limit. 

\section{Acknowledgments}
Computing resources
were provided by the 
University of Minnesota Supercomputing Institute.
We acknowledge the use of CMBFast.
MA, SH, and RS acknowledge support from 
NASA Grant NAG5-3941. 
JHPW and AHJ acknowledge support from 
NASA LTSA Grant no.\ NAG5-6552 and NSF KDI Grant no.\ 9872979. 
BR and CDW acknowledge support from NASA GSRP
Grants no.\ S00-GSRP-032 and S00-GSRP-031.
EFB acknowledges support from NSF grant AST-0098048.
The work of KAO was supported partly by DOE grant
DE--FG02--94ER--40823.
PGF acknowledges support from the Royal Society.
MAXIMA is supported by NASA Grant
NAG5-4454.

\appendix

\section{Probability Distribution of the Experimental Data}
\label{prob}
BJK have shown that the band-powers are well-approximated by an
offset log-normal distribution.  Specifically, they 
showed that the probability distribution $p(Z_i^{d}\ | \
Z_i^{t})$ is approximately Gaussian as a function of $Z_i^{t}$.
Furthermore, it is possible to compute the covariance matrix of
these Gaussian random variables.

The calculation in BJK was performed in a Bayesian framework.
For the frequentist analysis we need to
know the probability distribution $p(Z_i^{d}\ | \ Z_i^{t})$ as a
function of $Z_i^{d}$, not as a function of $Z_i^{t}$.  (This is the
heart of the difference between the two approaches: for a Bayesian the
data are fixed and the theoretical quantities are described
probabilistically; a frequentist treats the data as a random variable
for fixed values of the parameters.)  We therefore make the {\it
ansatz} that the probability distribution is Gaussian in $Z_i^{d}$ as
well.
\begin{table}
\begin{centering}
\begin{tabular}{c|c|c|}
    & Small Map &  Large Map  \\ \hline
${\bf\sf M}_{00}$&$1.5\times10^7\pm1.2\times10^6$&
$1.2\times10^8\pm2.8{\bf\times}10^6$
\\ \hline
${\bf\sf M}_{01}$&$2.0\times10^6\pm7.4\times10^4$&
$4.5\times10^6\pm7.9\times10^4$
\\ \hline
${\bf\sf M}_{11}$&$2.1\times10^7\pm2.6\times10^6$&
$1.2\times10^8\pm7.3\times10^6$
\\ \hline
\end{tabular}
\label{table:map}
\caption{ \footnotesize The average values
with sample standard deviations of the 
marginalized $Z_l$ weight matrix entries
for both large and small map simulations.  
The large map simulations
were based on pointing from the MAXIMA-1 $8'$ map, and 
the small map pointing was based
on a center patch of the map.  Units are
dimensionless MADCAP units}
\end{centering}
\end{table}

If the $Z_i^{d}$ are indeed Gaussian distributed, then the entries of the
weight matrix ${\bf\sf{M}}$ (inverse covariance matrix) should be exactly the
same for independent observations of universes which have the same
underlying CMB power spectrum. We test the assumption of Gaussianity using
simulations.  We generate CMB maps using a particular cosmological model,
compute the power spectrum and ${\bf\sf M}$ for each map and assess the
variance in the entries of ${\bf\sf M}$ between simulations. A small variance
would indicate that the assumption that the $Z_i^{d}$ are
Gaussian-distributed is adequate.

We generated 100 small-area and 30 large-area map simulations using
${\bf a}_0$ as the cosmological model and computed the ${\bf\sf M}$
matrix for each map (the number of simulations is limited by the
computational resources required to estimate the power spectrum for
each map).  The small- and large-area maps contain 542 and 5972 $8'$
pixels respectively.  Power spectra and $Z_l$ were computed in four
bins of $l=$\{2,300\},\{301,600\},\{601,900\},\{901,1500\}, and
${\bf\sf M}$ was obtained by marginalizing over the first and last bins.
The results are summarized in Table~\ref{table:map}1.  The average
value of the diagonal entries increases for the larger-area maps
because for those the band powers have smaller errors and hence larger
values in the weight matrix.  The percent fluctuation in the matrix
entries of the large-area maps are 2\% for the first diagonal entry,
6\% for the second diagonal entry, and 2\% for the off diagonal
entries.  We consider this variance to be small enough to indicate
that the assumption of Gaussianity of the $Z_i^{d}$ is acceptable.
We also note that the variance of the matrix elements decreases as a
function of increasing map size. If such a trend continues to maps of the
size of the MAXIMA-1 map, which has more than 15,000 pixels, then the
assumption of Gaussianity is well satisfied.

\bsp

\end{document}